# Semiregular tessellation of electronic lattices in untwisted bilayer graphene under anisotropic strain gradients

Zeyu Liu(刘泽宇)[1,2,3,#], Xianghua Kong(孔祥华)[1,#,*], Zhidan Li(李志聃)[1], Zewen Wu(吴泽文)[1], Linwei Zhou(周霖蔚)[1], Cong Wang(王聪)[2,3], and Wei Ji(季威)[2,3,*]

[1]*College of Physics and Optoelectronic Engineering, Shenzhen University, Shenzhen 518060, China.*

[2]*Beijing Key Laboratory of Optoelectronic Functional Materials & Micro-nano Devices, Department of Physics, Renmin University of China, Beijing 100872, China*

[3]*Key Laboratory of Quantum State Construction and Manipulation (Ministry of Education), Renmin University of China, Beijing, 100872, China*

*Emails: kongxianghuaphysics@szu.edu.cn (X.K.), wji@ruc.edu.cn (W.J.),

**ABSTRACT:** Two-dimensional (2D) moiré superlattices have emerged as a versatile platform for uncovering exotic quantum phases, many of which arise in bilayer systems exhibiting Archimedean tessellation patterns such as triangular, hexagonal, and kagome lattices. Here, we propose a strategy to engineer semiregular tessellation patterns in untwisted bilayer graphene by applying anisotropic epitaxial tensile strain (AETS) along crystallographic directions. Through force-field and first-principles calculations, we demonstrate that AETS can induce a rich variety of semiregular tessellation geometries, including truncated hextille, prismatic pentagon, and brick-phase arrangements. The characteristic electronic bands (Dirac and flat bands) of the lattice models associated with these semiregular tessellations are observed near the Fermi level, arising from interlayer interactions generated by the redistribution of specific stacking registries (AB, BA, and SP). Furthermore, the electronic kagome, distorted Lieb, brick-like, and one-dimensional stripe lattices captured in real-space confirm the tunable nature of the semiregular tessellation lattices enabled by AETS. Our study identifies AETS as a promising new degree of freedom in moiré engineering, offering a reproducible and scalable platform for exploring exotic electronic lattices in moiré systems.

# 1. Introduction

Two-dimensional (2D) moiré bilayers can be formed by stacking atomically thin van der Waals monolayers at controlled twist angles or through lattice mismatch. Localized interlayer electronic states arising from specific stacking registries act as artificial atoms, or termed "superatoms", arranging themselves into moiré superlattices that either preserve the symmetry of the parent atomic lattice or exhibit emergent symmetries beyond it. The inter-superatomic distances in those moiré lattices typically exceed those in conventional atomic crystals, substantially reducing electron kinetic energy between neighboring superatomic sites and resulting in narrower electronic bandwidths. Further suppression of kinetic energy occurs through rapid variations in local stacking configurations, leading to the formation of electronic flat bands. Such reduced kinetic energy magnifies electron-electron interactions, particularly electron correlation effects at the nanoscale, thereby enabling exotic quantum phases that are inaccessible in their bulk or pristine mono- and bi-layer counterparts[1, 2]. Extensive research has demonstrated these phenomena primarily in graphene[3-7], h-BN[8, 9], or transition metal dichalcogenides (TMDs)[10-15], which have already demonstrated diverse emergent states and electron correlation effects. Well known examples include correlated insulator[3-5, 8], superconducting[4, 8], fractional Chern insulator[7, 15], and the quantum anomalous Hall effect[13, 14].

Nearly all studied 2D moiré superlattices demonstrate regular tessellation lattice patterns, such as triangular, square, and hexagonal, lattices, in terms of both geometrical and electronic structures. Electron-electron interactions can further transform these regular lattices into more complex arrangements, such as kagome lattices, which is a type of semiregular tessellation (also called Archimedean tessellation) patterns that host characteristic electronic flat and Dirac bands[16, 17]. This naturally raises an intriguing question: can other semiregular tessellation lattices beyond the kagome lattice be realized, even only electronically, within bilayer moiré superlattices? Strain engineering has already proven effective as an additional in-situ tool for tuning electronic properties in both lattice-mismatched hetero-bilayers and twisted homo-bilayers[2, 18-20]. When combined with twist angles or intrinsic lattice mismatches, in-plane strain substantially expands the accessible property space moiré bilayers. Recent theoretical studies predict that isotropic biaxial strains applied differently to two untwisted graphene monolayers can induce kagome-like geometries upon structural relaxations, consequently yielding strain-dependent electronic states resembling those of kagome lattices[19, 21]. However, isotropic biaxial strain has been limited to kagome-like structures, leaving other semiregular tessellation lattices unexplored.

In this work, we propose and demonstrate additional semiregular tessellation electronic lattices in 2D, which are achievable by introducing anisotropic epitaxial tensile strain (AETS) into untwisted bilayer graphene (BLG). Force-field-based structural relaxations and density functional theory (DFT) calculations were used to reveal atomic and electronic structures in these moiré bilayers. Under various in-plane strain conditions, our calculations uncover several semiregular tessellation lattices beyond the kagome lattice. These lattices are characterized through unique interlayer spacing distributions, including truncated hextille, prismatic pentagon, and brick-phase arrangements. These geometrical arrangements are also associated with their characteristic electronic bands revealed from corresponding tight-binding models. Characteristic Dirac and flat bands near the Fermi level were confirmed using DFT-calculated electronic band structures and spatial distributions of corresponding electronic states.

## 2. Computational method and model

All structural relaxations were performed using the implementation of a force field, facilitated by the Large Atomic/Molecular Massively Parallel Simulator (LAMMPS).[22] The C-C interactions within each graphene layer were described using a second-generation reactive empirical bond order (REBO) potential,[23] while the interlayer van der Waals interactions were accounted for via the Kolmogorov-Crespi (KC) potential.[24] All atoms were fully relaxed until the residual force per atom was less than $1.0 \times 10^{-4}$ eV/Å. The lattice constant of the graphene primitive cell calculated by LAMMPS is 2.46 Å, and the initial non-strained configuration is Bernal stacked graphene. Density functional theory calculations were performed using the generalized gradient approximation (GGA) with the Perdew-Burke-Ernzerhof (PBE) for the exchange-correlation potential, linear combined atomic orbitals (LCAO) method, and a single-zeta polarized (SZP) atomic orbital basis set as implemented in the RESCU package.[25] The real space mesh resolution was set to 0.35 Bohr, and the convergence criteria for electronic energy and charge density are both set to $10^{-5}$ to ensure good convergence. The $k$-meshes of 3 × 3 × 1 were adopted for calculations. The models have a vacuum thickness up to about 26 Å in the z-direction to avoid interactions due to periodicity at the surface.

Herein, an anisotropic gradient-strained bilayer graphene (gs-BLG) moiré superlattice is constructed from a $3N$-1 ($a_1$) × $3N$ ($a_2$) graphene monolayer supercell and a bottom graphene monolayer with different period sizes. We theoretically align the axes of the two superlattices with different lattice constants, realizing an in-plane epitaxial strain gradient between the two layers in untwisted BLG, thereby forming a moiré pattern. We start from the reported SS-gs-BLG with two layers simultaneously subjected to isotropic tensile strains,[21] by gradually decreasing the strain freedom of the lattice axis of each layer and making the both layers share a certain superlattice constant.

The in-plane AES of the two layers can take four strategies, determined by the largest lattice constant of supercell axis: 1) as depicted in Fig. 1(a), when the two lattice axes of the top-layer are still subject to anisotropic tensile strain and only one lattice axis of the bottom-layer is subject to tensile strain, i.e., the superlattice period of the bottom-layer is $3N × 3N+1$, which we call this strained strategy "AETS-2+1-BLG"; 2) continuing to decrease the strain degrees of freedom of lattice axis strain in the bottom-layer (AETS-2+0-BLG), i.e., the superlattice period of bottom-layer is $3N+1 × 3N+1$ [Fig. 1(b)]; 3) the top-layer $a_1$ lattice axis and bottom-layer $a_2$ lattice axis are subjected to tensile strain (AETS-1+1-BLG), i.e., the superlattice period of bottom-layer is $3N × 3N$-1 [Fig. 1(c)]; 4) only one lattice axis of one layer is subjected to tensile strain (AETS-1+0-BLG), i.e., the superlattice period of bottom-layer is $3N × 3N$ [Fig. 1(d)]. In addition, if compressive strain is introduced, there will be a total of 16 cases (Table S1), and here we only consider the case of tensile strain.

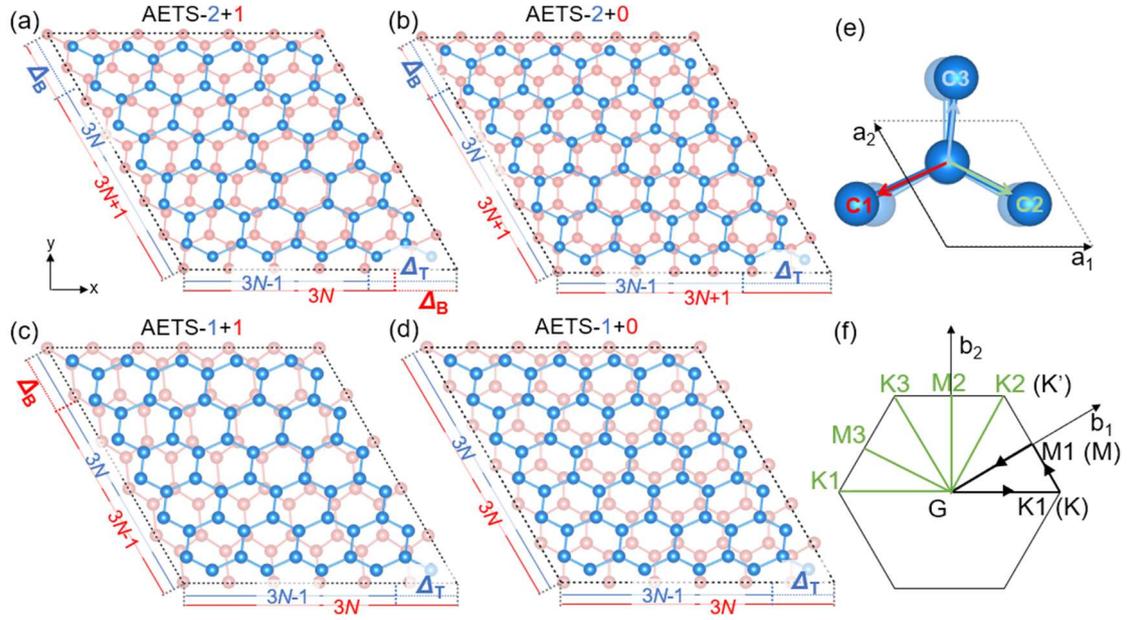

**Fig. 1.** Schematic top views of the models for (a) AETS-2+1-, (b) AETS-2+0-, (c) AETS-1+1- and (d) AETS-1+0- BLGs, where the top-layer (bule) and bottom-layer (red) have graphene supercells in different size. The top- or bottom-layers are subjected to potential strains to ensure that they have the same in-plane lattice constants in the bilayer supercell. The supercell is marked by the black dashed diamond. (e) Graphene cells subjected to AETS-1+0 with anisotropic variations in bond lengths and bond angles. (f) The first Brillouin zone of the AETS-BLGs, with black arrows showing the intrinsic high symmetry path of pristine graphene.

It is still important to emphasize here that the AETS is applied along the base-vector direction of the intrinsic hexagonal lattice of graphene, rather than the simplified orthogonal lattice. As a result, the bond angles and bond lengths formed by the C atom bonded to the three nearest atoms (C1, C2, C3) are anisotropic in Cartesian space, as shown in Fig. 1(e). This AETS along the crystallographic direction completely breaks the fundamental rotational and mirror symmetries of monolayer graphene, while new symmetries may emerge at larger moiré periods. The localized symmetry breaking leads to a change in the intrinsic Brillouin zone of graphene, which is insufficient to describe the electronic structure forming the global moiré superlattice, as shown in Fig. 1(f), where the Brillouin zone expands from the original black arrow path to half of the first Brillouin zone.

## 3. Geometric and electronic properties of $N=8$ AETS-BLGs

Building upon the $N=8$ top-layer graphene superlattice, we traverse 16 different types of AES-BLGs structures, which exhibit a rich variety of moiré patterns and stacking orders (Fig. S1). Interestingly, as shown in Fig. S2, AES-BLGs that do not involve compressive strain can maintain interlayer spacing within the conventional range for bilayer graphene (3.4 - 3.6 Å). However, the incorporation of compressive strain leads to the maximum interlayer spacing increases significantly to about 9.6 Å with significant out-of-plane corrugations, a phenomenon corroborated by experimental observations.[26] It is easy to understand that tensile strain provides atoms with additional space to adopt lower-energy stacking configurations, thereby effectively reducing both intralayer elastic energy and interlayer interaction energy without necessitating out-of-plane relaxation. Conversely, compressive strain constricts the in-plane lattice dimensions, compelling atoms to displace out of the plane to equilibrate the energy competition within and between layers, thereby minimizing the total energy of the system.

### 3.1 Geometric structure of AETS-BLGs

Figures 2(a-d) show the fully relaxed atomic structure of the four AETS-BLGs for $N=8$, displaying distinct values of in-plane strains: 8.70% and 4.17% tensile strain for the top-layer along the $a_1$ and $a_2$, 4.17% tensile strain for the bottom-layer along the $a_1$ in AETS-2+0-BLG [case 1, Fig. 2(a)], 8.70% and 4.17% tensile strain for the top-layer along the $a_1$ and $a_2$ in AETS-2+0-BLG [case 2, Fig. 2(b)], 4.35% tensile strain for the top-layer and bottom-layer in AETS-1+1-BLG [case 3, Fig. 2(c)], and 4.35% tensile strain for the top-layer in AETS-1+0-BLG [case 4, Fig. 2(d)].

AETS-BLG moiré superlattices contain six different high-symmetry stacking configurations, namely AA, AB, BA, SP1, SP2, and SP3, As shown in the right inset of Fig. 2(d). In the AA stacking (blue in Fig. 2), each atom in the top-layer is directly above a carbon atom in the bottom-layer, and the lateral movement of half-units produces either the AB (purple) or the BA (orange) stacking. Furthermore, the introduction of anisotropic epitaxial strains disrupts the intrinsic $C_3$ symmetry of the graphene layer, lifting the degeneracy of the original transition SP (saddle-point) stacking (green), and forming three distinct stacking configurations: SP1 (short dashed line), SP2 (long dashed line) and SP3 (solid line). Relaxed AETS-2+1-, 2+0- and 1+1-BLGs structures feature patterns combining deformed hexagonal (AB and BA regions, in violet and orange) and triangular (AA regions, in blue) lattices, as depicted in Figs. 2(a-c). However, the AETS-1+0-BLG shows a stripe pattern spaced by AB, SP2, BA, SP1 regions in Fig. 2(d). Since the direction of epitaxial strain is along the zigzag direction, the AETS-1+0-BLG does not appear AA and SP3 stackings.

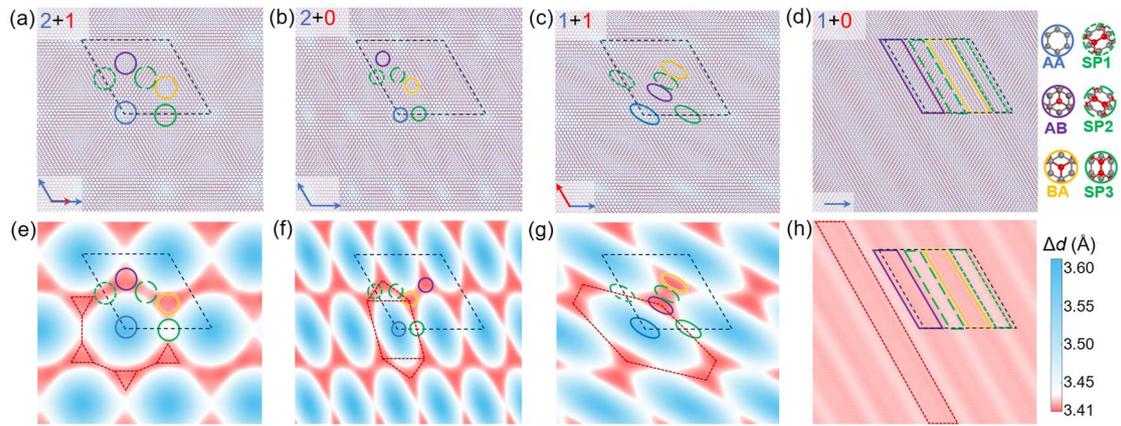

**Fig. 2.** Geometries of AETS-BLGs. (a) In-plane moiré patterns of AETS-2+1-, (b) 2+0-, (c) 1+1-, (d) 1+0-BLGs. Blue and red arrows in the lower left corner represent the top- and bottom-layers subjected to strain directions. (e-h) Projections of the corresponding interlayer spacing of the four. The supercell is marked by the black dashed diamond, and the white in color bar corresponds to the optimal interlayer spacing of the non-strained SP stacking. Within the diamond, six high-symmetry stacking orders are denoted by differently colored circles, and atomic models on the right side of the (d) are presented for these six stacking orders inside the circles with associated colors.

Different in-plane strain combinations will significantly reshape the spatial periodicity of stacking orders, leading to the emergence of out-of-plane corrugation, thereby forming distinct interlayer spacing ($\Delta d$) distribution patterns. Figures. 2(e-h) show the interlayer spacing maps of the four AETS-BLGs, highlighting the relative out-of-plane corrugations. The white region is the optimal interlayer spacing for non-strain SP stacking, which distinguishes the AA and AB/BA stacking domains. In both AETS-2+1-BLG and AETS-2+0-BLG, SP1 and SP2 stackings remain continuous in the interlayer spacing maps, while SP3 is spatially separated. The latter results in periodic compression of the stacking order due to the absence of epitaxial strain along the $a_1$ direction in the bottom-layer. In AETS-1+1-BLG, SP1, SP2, and SP3 are all spatially separated, and the anisotropy of AB and BA stacking distances is stronger. As for AETS-1+0-BLG, the absence of AA stacking results in negligible out-of-plane structural corrugation.

Furthermore, under different AETS combinations, the strongly coupled AB and BA stacking domains form four novel semiregular tessellation patterns: truncated hextille [AETS-2+1-BLG, Fig. 2(e)], prismatic pentagon [AETS-2+0-BLG, Fig. 2(f)], brick-phase [AETS-1+1-BLG, Fig. 2(g)] and 1D stripe [AETS-1+0-BLG, Fig. 2(h)] arrangements, delineated by red dashed triangles and lines. The formation of in-plane stacking orders enriches the landscape of local interlayer interactions, while the competition between in-plane elastic (strain) energy and interlayer coupling energy drives the reconstruction of interlayer-spacing patterns. Specifically, AA stacking regions reduce the system energy by increasing the interlayer spacing to approximately 3.6 Å, whereas AB/BA stacking domains maintain interlayer distances close to that of intrinsic Bernal stacking in graphite (~3.4 Å), in agreement with previously reported observations.[27, 28]

## 3.2 Electronic properties in AETS-BLGs

The various semiregular tessellation lattices observed in the layer-spacing projection patterns [Figs. 2(e-h)] can host exotic electronic states, whose eigenbands, particularly the characteristic flat and Dirac bands, closely relate to the electronic bandstructures predicted by tight-binding models of the corresponding lattices.[17, 29-31] Such distinctive band features are anticipated to emerge prominently in anisotropic strain gradient-engineered untwisted bilayer graphene (AETS-BLG), a prediction explicitly verified by our DFT calculations. Figure 3(a) illustrates the layer-projected band structure of

AETS-2+1-BLG, highlighting two sets of kagome-derived bands featuring Dirac and flat bands. The green kagome bands comprises a Dirac point (DP1) and a higher-energy flat band (FB1), primarily originating from the top-layer. In contrast, the brown kagome band includes a lower-energy flat band (FB2) and another Dirac point (DP2), mainly contributed by the bottom-layer. Notably, the Dirac points open energy gaps due to the breaking of $C_3$ symmetry [Fig. S3(a, b)], and the two flat bands exhibit opposite signs of the effective hopping parameters. For AETS-2+0-BLG, similar two sets of Dirac bands (DP1/DP2) and flat bands (FB1/FB2) are also observed near the Fermi level, as shown in Fig. 3(b). The symmetry of the non-strained bottom-layer remains intact, so the Dirac point remains closed. The degeneracy of the flat band and the Dirac bands at the G point is broken, with the flat band gradually moving towards the position of the Dirac point. This reminds us of the band structure evolution of the transition from the kagome lattice to the Lieb lattice, associated with the lattice symmetry. [30]

Figure 3(c) presents the band structures of AETS-1+1-BLG, which reveals Dirac bands and two flat bands near the Fermi level (highlighted in green), where the Dirac bands originate from the top-layer and the flat bands are contributed by both layers. Due to the different strain directions in the top- and bottom-layers, similar band features are present in G-K3 high symmetry path, where the Dirac bands are contributed by the bottom-layer [Fig. S4(c)]. Interestingly, while epitaxial strain alone can slightly shift Dirac point and open energy gap in monolayer graphene [Fig. S3(e, h)], the formation of the moiré superlattice introduces a brick-like lattice composed of AB/BA domains, which restores a new mirror symmetry. The resulting interlayer coupling leads to the closure of the Dirac energy gap. For AETS-1+0-BLG, which is structurally composed of strongly interlayer-coupled AB, BA, and SP stackings, the Dirac bands near the Fermi level degenerate into parabolic shapes and gradually approach each other, resulting energy band crossings, as marked in green in Fig. 3d. The interlayer interactions induced by the 1D stripe-like moiré superlattice enhance anisotropy in the band structure along different momentum paths, such as the Dirac bands flatten along the K-M direction.

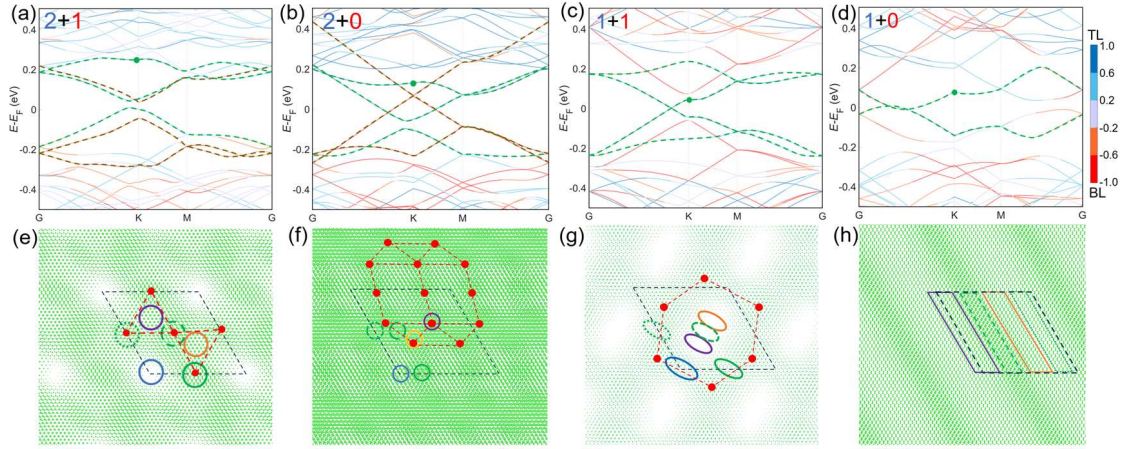

**Fig. 3.** Layer atomic projection energy band diagrams for (a) AETS-2+1-, (b) AETS-2+0- (c) AETS-1+1-, and (d) AETS-1+0-BLGs. Gray-blue colored lines represent bands combinedly contributed from both the top-layer and bottom-layer, while the light blue and orange lines indicate dominance by the top-layer and bottom-layer. Darker blue (red) lines, signify nearly exclusive contribution from the top- (bottom) layer. (e) The interlayer $|\varphi|^2$ of AETS-2+1-BLG for the flat band (green dot) in (a), where the isosurface value of the wavefunctions are $3\times10^{-5}$ e Å$^{-3}$. (f) The interlayer $|\varphi|^2$ of AETS-2+0-BLG for the flat band (green dot) in (b), where the isosurface value of the wavefunctions are $1\times10^{-5}$ e Å$^{-3}$. (g) The interlayer $|\varphi|^2$ of AETS-1+1-BLG for the Dirac band (green dot) in (c), where the isosurface value of the wavefunctions are $6\times10^{-5}$ e Å$^{-3}$. (h) The interlayer $|\varphi|^2$ of AETS-1+0-BLG for the Dirac band (green dot) in (d), where the isosurface value of the wavefunctions are $3\times10^{-5}$ e Å$^{-3}$. Black dashed diamonds and differently colored circles represent the supercell and the six stacking orders in the same scheme used in Fig. 2.

The interlayer real-space wavefunction norm squared $|\varphi|^2$ further reveals the intimate connection between these signature energy bands and their underlying lattice models. Figure 3(e) displays $|\varphi|^2$ of FB1 at the high-symmetry point K [marked in Fig. 3(a)], showing strong localization in the SP1, SP2, and SP3 stacking regions, forming an electronic kagome lattice, further confirming the presence of kagome flat bands in AETS-2+1-BLG. In contrast, for AETS-2+0-BLG, the regions where the $|\varphi|^2$ of FB1 accumulates (AB, BA, and SP3) connect to form a prismatic pentagon lattice in real space, similar to the pattern observed in the interlayer-spacing projection, resembling a distorted Lieb lattice. This is consistent with the shift of the flat band position observed in the electronic bandstructure, indicating a strain-tunable transition from an electronic kagome lattice to an electronic distorted Lieb lattice (Fig. S5). Figure 3(g) presents the $|\varphi|^2$ of the Dirac band at the K point. Unlike the previous two cases, the $|\varphi|^2$ aggregation region is not in specific high-symmetry stacking regions. Instead, it is localized at the edge regions outside the AB, BA, and SP2 stacking domains (marked by red dots), forming a brick-phase lattice. Together with the characteristic bands observed in the bandstructure, this further confirms the emergence of an interlayer electronic brick-like lattice modulated by epitaxial strain. In the 1D-striped moiré superlattice formed in ATES-1+0-BLG, the interlayer Dirac electronic states are modulated by the striped periodic potential and are confined within the SP2 stacking regions, forming a stripe lattice, as shown in Fig. 3(h).

## 4. Conclusion

In summary, we have demonstrated that anisotropic epitaxial strain gradients provide a powerful and tunable route for engineering semiregular tessellation of electronic lattices in 2D moiré bilayers. Employing fully atomistic relaxations based on force-field calculations, we reveal that in-plane AETS efficiently induces the redistribution of the interlayer stacking orders, leading to diverse semiregular tessellation patterns, such as truncated hextille, prismatic pentagon, and brick-phase arrangements. Through first-principles calculations, we further reveal that beyond the kagome electronic lattice structures previously observed under isotropic biaxial strain, distinctive electronic bands corresponding to distorted Lieb and brick-wall lattices also emerge near the Fermi level. These special lattice band models correspond to the semiregular tessellation patterns observed in interlayer-spacing projections, stemming from the spatial redistribution of strongly coupled AB/BA and SP stacking domains, which give rise to distinct interlayer interactions. The electronic semiregular tessellation lattices, composed of localized interlayer electronic states arising from specific stacking registrations, further corroborate the topological lattices formed by the semiregular tessellation patterns. These results thus validate AETS as a viable strategy for engineering the emergence of semiregularly tessellated electronic lattices. Additionally, we find that the prismatic pentagon pattern observed in the interlayer-spacing projection of AETS-2+0-BLG yields electronic band structures featuring signatures of distorted Lieb lattices, indicating AETS as an effective means to drive the evolution from kagome to Lieb lattice structures. Collectively, this study establishes anisotropic epitaxial strain as a generalizable degree of freedom in moiré materials design, offering a novel pathway for exploring other (semi-)regular tessellation lattices.

**Acknowledgment**

We thank Drs. Kui Gong, Yibin Hu, and Yin Wang (all from HZWTECH) and Prof. Yiqi Zhang for helpful discussions. We gratefully acknowledge the financial support from the National Natural Science Foundation of China (Grants No. 52461160327, No. 92477205, No. 12474173 and No. 12104313), the National Key R&D Program of China (Grant No. 2023YFA1406500), the Department of Science and Technology of Guangdong Province (No. 2021QN02L820) and Shenzhen Science and Technology Program (Grant No. RCYX20231211090126026, the Stable Support Plan Program 20220810161616001), the Fundamental Research Funds for the Central Universities, and the Research Funds of Renmin University of China (Grants No. 22XNKJ30). Calculations were performed at the Physics Lab of High-Performance Computing (PLHPC) and the Public Computing Cloud (PCC) of Renmin University of China.

*B* **102** 035142